\documentstyle[twocolumn,aps,prl,graphicx,amssymb]{revtex}

\begin{document}

\draft

\title{Synchronization learning of coupled chaotic maps}

\author{Luis G. Moyano$^1$\thanks{E-mail address:
moyanol@cab.cnea.gov.ar},
Guillermo Abramson$^{1,2}$\thanks{E-mail address:
abramson@cab.cnea.gov.ar},
and Dami{\'a}n H. Zanette$^{1,2}$\thanks{E-mail address:
zanette@cab.cnea.gov.ar}}
\address{$^1$Centro At{\'o}mico Bariloche and Instituto Balseiro, 8400
Bariloche, Argentina \\
$^2$Consejo Nacional de Investigaciones Cient\'{\i}ficas y T\'ecnicas,
Argentina}

\date{\today}

\maketitle

\begin{abstract}

We study the dynamics of an ensemble of globally coupled chaotic
logistic maps under the action of a learning algorithm aimed at
driving the system from  incoherent collective evolution to a
state of spontaneous full synchronization. Numerical
calculations  reveal a sharp transition  between regimes of
unsuccessful   and   successful learning as   the algorithm
stiffness grows. In the  regime of successful learning, an
optimal value of the stiffness is found for which the learning
time is minimal.

\end{abstract}

\pacs{PACS numbers: 05.45.Xt, 05.45.-a, 87.23.Kg}

\section{Introduction}

Synchronization is a form of macroscopic  evolution observed in a wide
class of complex systems. Typically, it appears when  the range of the
interactions inside the  system is of same order  as  the system size.
Mechanical  and  electronic  devices,  as   well  as certain  chemical
reactions \cite{Kura} are known  to exhibit synchronized dynamics.  In
the   realm   of biology,   among   many   other instances   \cite{W},
synchronization appears at    the cellular level   in neural  networks
\cite{NN} and  in heart tissues  \cite{heart}. Animal populations show
also  complex forms   of synchronous  behavior,  the  most spectacular
example being probably the  synchronous flashing of  certain fireflies
\cite{ff}. The spontaneous occurrence of synchronization in biological
systems suggests  that this form of collective  behavior develops as a
consequence  of evolutionary  selection or of   some kind  of adaptive
learning.  In this paper we   explore a model  where  a set of coupled
chaotic elements is added with a feedback learning process targeting a
collective    synchronized   state.   We find    that, under  suitable
conditions, the system evolves  from completely incoherent behavior to
a state of full synchronization.

Collective behavior under the effect  of long-ranged
interactions  can be  modelled  by means  of    ensembles of
globally  coupled  dynamical systems. Introduced   by Kaneko a
decade  ago \cite{kaneko}, globally coupled  logistic maps have
proven  to be an  appropriate paradigm for such kind of emerging
evolution. The  system consists of $N$ identical mappings whose
individual dynamics, in the absence of interactions, is given by
$x(t+1)=F[x(t)]$ with $F(x)=  r   x(1-x)$.  The individual
dynamics are coupled according to
\begin{equation}
x_{i}(t+1)=(1-\epsilon)F[x_{i}(t)]+\frac{\epsilon}{N}\sum_{j=1}
^{N}F[x_{j}(t)],
\label{Kaneko}
\end{equation}
($i=1,\dots,N$), where $\epsilon \in (0,1)$ is the coupling
intensity.

The  system  defined  by Eqs. (\ref{Kaneko})   features  a  variety of
collective behaviors in the space  spanned by the nonlinear  parameter
$r$ and the    coupling constant $\epsilon$   \cite{kaneko}.  At large
values of $\epsilon$ the ensemble is in a {\it coherence phase}, where
all the  elements tend asymptotically  to  exactly the same trajectory
$x(t)$.  For sufficiently    long times,  thus,  the system   is fully
synchronized. Note that the trajectory of the synchronized ensemble is
governed by the dynamics of a single element.  At low $\epsilon$, when
coupling  is  weak, a  {\it  turbulence phase} is observed,  where the
evolution is completely unsynchronized. At  intermediate values of the
coupling  intensity a {\it clustering  phase} with different groups of
mutually synchronized elements appears.

Bearing in mind the role of synchronization as a collective
acquired behavior of biological systems, we incorporate to model
(\ref{Kaneko}) an additional evolutionary mechanism. Concretely,
each element is allowed to vary  its coupling constant  in time,
so that   the effect of  the collective evolution on its
dynamics---defined by the last term in the right-hand side of
(\ref{Kaneko})---is modified according to  a given criterion.
Thus, the system may be able to learn to perform a specific
collective task, in particular to evolve towards a coherent
synchronized state. We focus the attention in the properties of
the learning process, and find that a regime of successful
learning exists for a sufficiently stiff algorithm. Within this
regime, the time necessary to achieve synchronization is
determined by the stiffness.

\section{Modeling}

\label{modeling}

We  consider a  variation  of model (\ref{Kaneko})  where the coupling
constants depend  on   time and  may be  different   for each element,
namely,
\begin{equation}
x_{i}(t+1)=[1-\epsilon_i(t)] F[x_{i}(t)]+\frac{\epsilon_i(t)}{N}
\sum_{j=1} ^{N}F[x_{j}(t)].
\label{K1}
\end{equation}
Inhomogeneous  global coupling in   ensembles  of logistic  maps  with
time-independent coupling  constants  has been considered in  previous
work \cite{inh1,inh2}. It can  be shown that  the system exhibits full
synchronization if and only if all the coupling constants $\epsilon_i$
are larger than a certain value  $\epsilon_c$. This value turns out to
coincide with the critical point for the onset of full synchronization
in homogeneous ensembles,  and can be given  in terms  of the Lyapunov
exponent $\lambda$ of a  single map as  $\epsilon_c=1-\exp (-\lambda)$
\cite{Lyap}.  In particular, for  a  nonlinear parameter $r=4$ we have
$\lambda = \ln 2$ and therefore $\epsilon_c=1/2$.

Time variation of coupling intensities as  a form of adaptive
behavior has been considered in globally  coupled maps by analogy
with synaptic evolution in neural networks \cite{inh1}, and  in
models of asymmetric imitative dynamics  for   two-element
systems \cite{imit}.  Here,   we consider  a    learning algorithm
based on   a   comparison  of the instantaneous state $x_i(t)$ of
each element with a global property of the ensemble, namely the
instantaneous average    state $\langle x(t)\rangle =
N^{-1}\sum_j  x_j(t)$.   If  the  distance from the individual
state to the average is larger than a certain threshold $u$ the
learning process acts, and the coupling constant of the element
is changed  to a new value, chosen  at random from a uniform
distribution in $(0,1)$. Otherwise, $\epsilon_i$ remains
unchanged.  Explicitly,
\begin{equation} \label{learning}
\epsilon_i(t+1) = \left\{
\begin{array}{ll}
\xi_i(t+1) & \mbox{ if $|x_i(t+1)-\langle x(t+1)\rangle|>u$} \\
\epsilon_i(t) & \mbox{ otherwise},
\end{array}
\right.
\end{equation}
where   $\xi_i$   is a   random  number with   uniform
distribution in $(0,1)$. Note that the threshold $u$  can be
interpreted as an inverse measure of the  stiffness  of learning.
The    evolution  proceeds  according   to    the following
dynamical rules. First,    the state $x_i(t)$ of every    element
is  updated to $x_i(t+1)$ applying map   (\ref{K1}).   The
average state $\langle   x (t+1)  \rangle$ is  then calculated.
Finally,  the learning algorithm (\ref{learning})  is   applied
to  every   element. This  procedure is successively  iterated,
so that each  evolution  step consists of two substeps where
coupled dynamics  and learning act  sequentially. Both processes
are applied synchronously to the whole system.

This is a  form of stochastic unsupervised learning \cite{Haykin}
where the whole ensemble is expected to selforganize into a
coherent state where the orbit of  every element coincides with
the average trajectory. The algorithm can be interpreted as an
adaptive control mechanism \cite{control}. Recently, control
techniques have been proposed to drive both low-dimensional and
extended dynamical systems towards a prescribed state, such as a
particular spatiotemporal pattern \cite{ding,sinha}. The present
variant is inspired in arguments of biological plausibility.
First, the target of learning is not a specific dynamical state,
but a wide class defined by a collective property, namely
synchronization. This class includes not only infinitely many
synchronized orbits but also a variety of different
configurations of the set of couplings $\epsilon_i$. Second, the
learning algorithm acts at the individual level. That is, at each
time step the criterion (\ref{learning}) is applied to each
element. According to its individual state a modification is
introduced to its coupling. The collective state achieved through
learning emerges thus as a consequence of individual evolution.
Finally, the modification applied to the coupling intensities is
random and unbiased. No hints are given on the desired values of
the coupling intensities, which must be adaptively found by the
system through iterations of trials and errors. Successive values
of $\epsilon_i$ are completely uncorrelated.

In the following  section,  we report  results  of extensive
numerical realizations of the  above model.  They correspond  to
an ensemble of logistic   maps  with  $r=4$,  i.e. at    the
fully developed  chaotic regime. The  system  is investigated as
a  function of  the threshold $u$. We detect  a  sharp transition
at  $u=0.5$,  between a regime  of successful learning  and a
regime where learning  fails. The origin of this transition is
identified   by studying the  intermittent dynamics just before
the state of full  synchronization is reached. We define a
parameter   that measures the    performance of learning   and
find an optimal value of $u$ for which learning is fastest.

\begin{figure}
\centering \resizebox{\columnwidth}{!}{\includegraphics{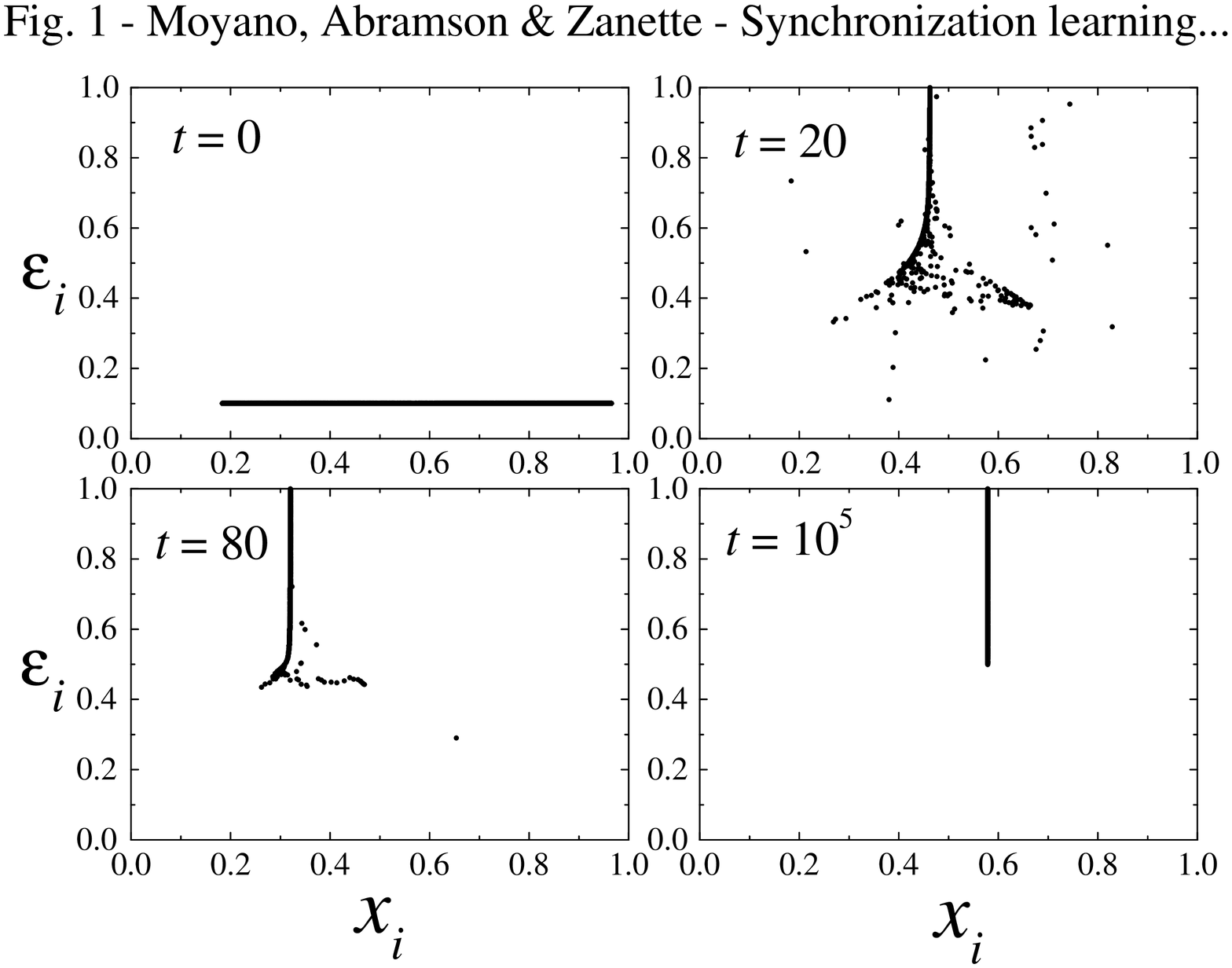}}
\caption{Snapshots of an   ensemble   of  $N=1000$  elements  on   the
$(x_i,\epsilon_i)$-plane at  four times:  $t=0$,  $t=20$,
$t=80$,  and $t=10^5$. The   learning   threshold is   $u=0.2$
and  learning   is successful.}
\label{evsx}
\end{figure}

\section{Results}

\label{results}

The numerical results reported in  this section correspond to
systems with $N=10^3$ elements. The initial states $x_i(0)$ are
distributed at random, with uniform density,     in the interval
$(0,1)$.   All   the coupling constants  have  initially  the
same   value, $\epsilon_i(0)= \epsilon_0=0.1$. For this coupling
intensity and a nonlinear parameter $r=4$, the initial state  of
the system is  well within the turbulence phase \cite{kaneko}.
During  a first stage, the  ensemble is  left to evolve $10^3$
steps without applying  the learning dynamics, such that the
individual states $x_i$ adopt the characteristic distribution of
an incoherent state. After  this,  time  is reset  to zero,
learning is switched on, and the states and coupling constants of
all elements are recorded during the following $10^4$ to $10^6$
steps.

A suitable  way of representing the instantaneous  state of the system
is a plot where each element is shown as a dot in the plane spanned by
the individual state   $x_i$  and the coupling   constant $\epsilon_i$
\cite{inh1,inh2}. Figure \ref{evsx}   shows four such snapshots for  a
single   system  at  different  times,  corresponding   to a threshold
$u=0.2$. At the initial time $t=0$ all elements have the same coupling
constant $\epsilon_0$, and  form   an incoherent, extended    cloud in
$x_i$. At subsequent times, $t=20$ and $t=80$, the effects of learning
are clearly visible. The elements that  have migrated to larger values
of $\epsilon_i$ form now a rather compact  cluster though, for $t=20$,
many elements with  relatively large coupling  constants are still far
from the main cluster. For  $t=80$, almost all  the elements have  got
coupling  constants above   the critical   value $\epsilon_c=0.5$. The
remaining elements   form a small  cloud  just below $\epsilon_c$, and
approximately follow   the motion of the  main  cluster.  Finally, for
$t=10^5$ all the elements have $\epsilon_i>0.5$ and  the same value of
$x_i$.  For this value  of $u$, learning has  been  successful and the
ensemble has become fully synchronized.

\begin{figure}
\centering \resizebox{\columnwidth}{!}{\includegraphics{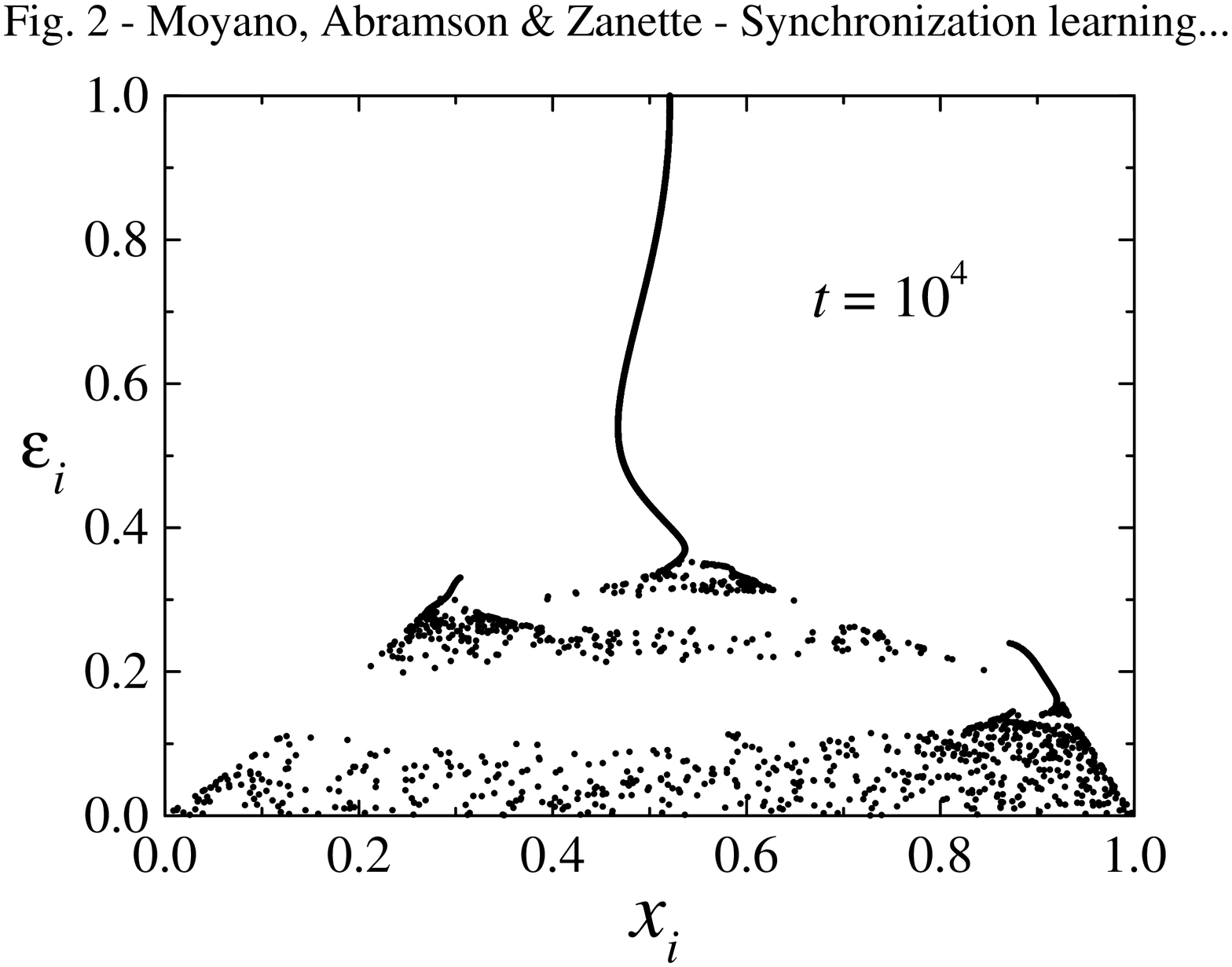}}
\caption{Snapshot   of  an  ensemble  of  $N=1000$   elements  on  the
$(x_i,\epsilon_i)$-plane  with $u=0.6$ at $t=10^4$.   At this
value of the  threshold  $u$ learning   fails  to  drive  the
system to   full synchronization.}
\label{evsx1}
\end{figure}

We show in Fig.   \ref{evsx1} a situation  where,  on the  other hand,
learning is  not  able  to lead   the system  to the   synchronization
phase.  Here $u=0.6$   and the system   has been  left  to evolve  for
$t=10^4$ steps.   After an  initial  redistribution   of the  coupling
constants, the  ensemble appears  to  have reached a stationary  state
with a complex organization  in the $(x_i,\epsilon_i)$-plane  (compare
with analogous distributions reported in \cite{inh1,inh2}) but with no
traces of synchronization. According to our simulations, this state is
preserved at longer times, of the order of $10^6$ steps.

These preliminary results suggest that some  kind of transition occurs
at   an intermediate  value $u_c$ of    the threshold, separating  two
regions     where   learning    is     respectively   successful   and
unsuccessful. This transition is characterized in the following.

\subsection{Learning transition}

As a global measure of the collective state of the ensemble during its
evolution we have chosen the mean  dispersion of the individual states
$x_i$,  namely,  $\sigma_x    =\sqrt{\langle  x^2   \rangle-\langle  x
\rangle^2}$. For the fully  synchronized state, $\sigma_x=0$. We  have
studied the  evolution of   $\sigma_x(t)$  for several values   of the
threshold     $u$,  and  found  that   for   $u>0.5  $  the dispersion
asymptotically  approaches  a   finite value.  The   average  of  this
asymptotic value over different realizations for  a fixed threshold is
practically independent of $u$, $\bar  \sigma_x \approx 0.25$. This is
the   dispersion   that corresponds       to the   state    shown   in
Fig. \ref{evsx1}.

For $u<0.5$, on the  other hand, we  have always found  that $\sigma_x
\to 0$ for   sufficiently long  times. We  stress,  however, that  the
typical times associated with this   evolution depend strongly on  the
threshold, as shown in detail  later. In any  case, for such values of
$u$, the   ensemble  approaches asymptotically   the state   of   full
synchronization.  Thus, the critical threshold $u_c\approx 0.5$ is the
boundary  between the zone  of  successful learning  ($u<u_c$) and the
zone where the system fails to learn how  to synchronize ($u>u_c$). As
discussed,  the  average   asymptotic  dispersion $\bar   \sigma_x$ is
sharply discontinuous at $u_c$.

\begin{figure}
\centering \resizebox{\columnwidth}{!}{\includegraphics{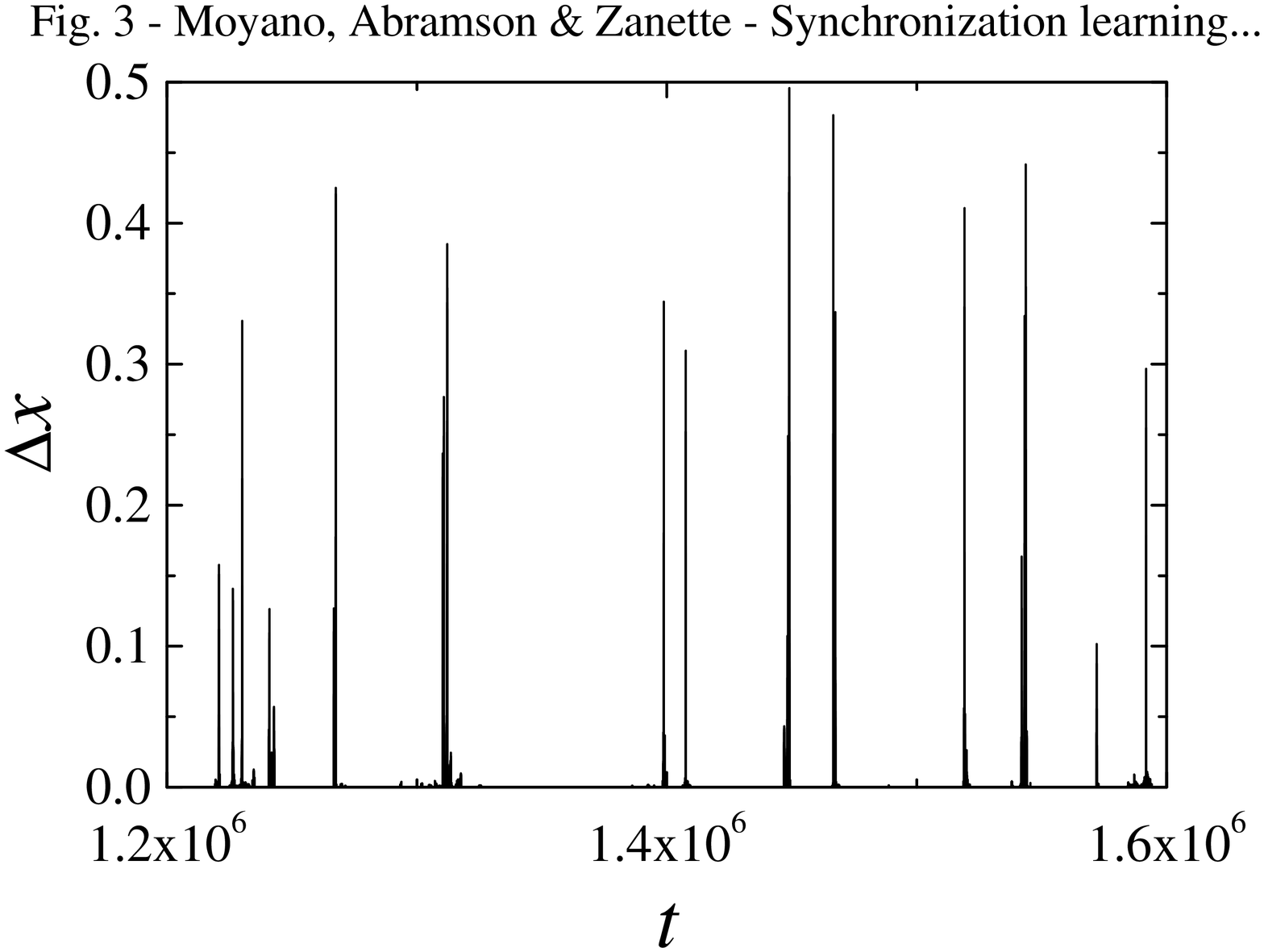}}
\caption{Intermittent evolution of  a single element that is prevented
to evolve, with coupling  constant $\epsilon_i=0.4997$. The plot
shows the absolute  difference between  the state of   the
element and  the average state of the system. The learning
threshold is $u=0.2$.}
\label{intermitent}
\end{figure}

Actually, this abrupt transition in the performance of learning
can be explained  by studying the dynamics  of our system in the
intermediate stages of  evolution, when  a  considerable fraction
of  the ensemble already  defines a  compact  cluster whereas the
remaining  elements, whose coupling constants are just  below
$\epsilon_c=0.5$,  form  the small cloud depicted in Fig.
\ref{evsx} for $t=80$. In this situation, the system is in the
threshold of a bifurcation where the stable state of full
synchronization appears. Indeed,  it would suffice to slightly
change the coupling  constants of the elements  in the cloud to
values above  $\epsilon_c$ in order  to  create an attractor
corresponding to the synchronized  state.  In the  threshold   of
such bifurcation  the system is expected to display intermittent
evolution \cite{interm}. We have in  fact verified in the
numerical simulations that the dynamics of each  element in the
cloud  exhibits two distinct regimes.  Most of the time,  the
element is found  in  a ``laminar'' regime,  where its state is
practically equal to that of  the main cluster. Occasionally,
however, the evolution  exhibits ``turbulent'' bursts during
which the element performs short excursions far away from the
cluster.  In order to illustrate  this   behavior, we  have
recorded the   evolution of a single element in the  ensemble,
whose coupling constant is  initially fixed at
$\epsilon_i=0.4997$.  Learning is subsequently prevented  for
this special element, in such a way that its coupling constant
remains fixed as   time elapses. For  long  times, $t\sim 10^6$,
the coupling constants   of all       the     other elements
are      found  above $\epsilon_c$.  Intermittency is  apparent
in  Fig. \ref{intermitent}, where the difference $\Delta x=|x_i -
\langle x \rangle |$ has been plotted for the  special element as
a function   of time. Here,  the average state $\langle x \rangle$
is essentially determined  by the position of the main cluster.

As expected,  the intervals of ``laminar''  behavior are found to grow
in length  as the   coupling  constant of   the  element  under  study
approaches $\epsilon_c$.   Conversely, the frequency  of ``turbulent''
bursts decreases.  Note that, for the  elements in the cloud, learning
is  possible  during these  bursts   only.  In   fact, according    to
(\ref{learning}),  learning acts   when the   difference  between  the
individual  state and the  average  state is  large enough. Thus,  the
closer the coupling constant is  to $\epsilon_c$, the later an element
undergoes a learning step.

The dynamics of  a single element  in the cloud below $\epsilon_c$ can
be  well  approximated as  follows.  We disregard   the  effect of the
remaining elements in  the cloud and suppose  that the  interaction of
the element under study with the ensemble occurs only through the main
cluster.   Conversely,  we  suppose that  the    main cluster contains
essentially all the   elements of the  system in  a synchronized state
$x_0(t)$,  and  that is not affected   by the dynamics   of the cloud.
Within these assumptions, the state $x(t)$ of the element under study,
whose  coupling constant is   $\epsilon \lesssim \epsilon_c$,  evolves
according to
\begin{equation} \label{1el}
x(t+1) = (1-\epsilon) F[x(t)] +\epsilon F[x_0(t)].
\end{equation}
Meanwhile,  the state of the main cluster obeys the dynamics of a
single independent map, $x_0(t+1)=F[x_0(t)]$.
The analytical study of the intermittent
evolution of $x(t)$ from Eq. (\ref{1el}) may be difficult. However, we
can easily find bounds for the excursions of $x(t)$ during the bursts.
Note in fact that we can write
\begin{eqnarray}
\Delta x(t+1) &=& |x(t+1)-x_0(t+1)|  \nonumber \\
&=& (1-\epsilon) | F[x(t)]-x_0(t+1) |.
\label{delta}
\end{eqnarray}
Putting $F(x)=4x (1-x)$ and taking into  account that at any time both
$x(t)$ and $x_0(t)$ are in  the interval $(0,1)$, we  find $ \Delta  x
<1-\epsilon$ which, for $\epsilon \to \epsilon_c$ reduces to
\begin{equation} \label{delta1}
\Delta x < 1/2.
\end{equation}
This bound is  clearly  seen in Fig. \ref{intermitent}.    We
conclude that  the distance from the main  cluster to an element
in the cloud during a  burst  cannot be larger  than $1/2$. For
such an element, consequently, learning will   occur during a
sufficiently  ample burst only if $u<0.5$. If $u>0.5$,    the
learning algorithm will never   be applied and the system will
not reach  the fully synchronized  state. This fixes the learning
transition at $u_c=0.5$.

\subsection{Learning times}

In order to give  a more detailed  description of the evolution of our
system under the action of learning, we focus now the attention on the
time needed to approach  the synchronization state.   To define such a
time, we  study  first  the fraction $n(t)$   of  elements that,  at a
certain moment, have  their coupling constants below $\epsilon_c$.  We
have $n(0)=1$ and, for $u<u_c$, $n(t) \to 0$ as $t\to \infty$.

\begin{figure}
\centering \resizebox{\columnwidth}{!}{\includegraphics{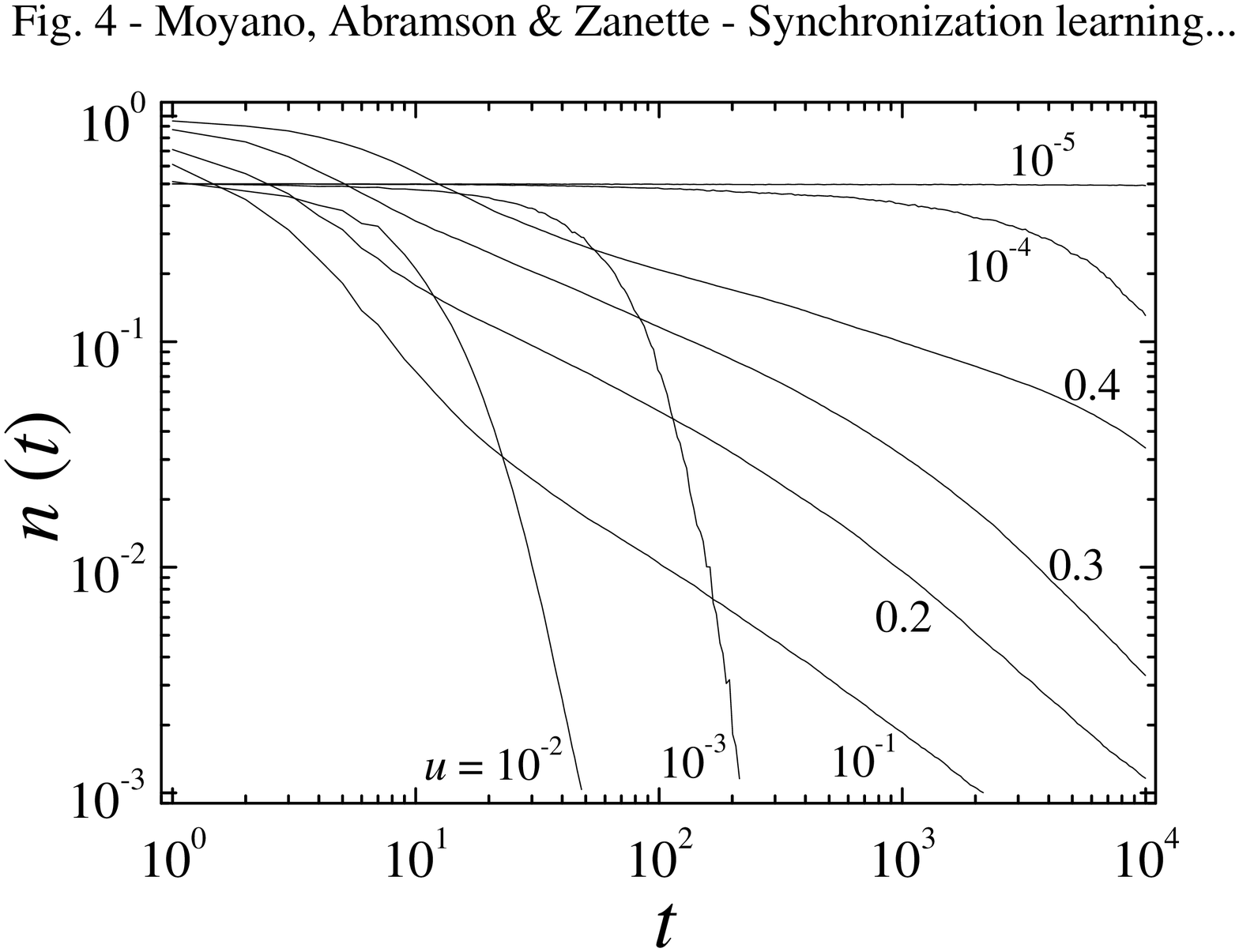}}
\caption{Time   evolution of the    fraction $n(t)$  of elements  with
$\epsilon_i(t)<0.5$ in   ensembles  of  $N=1000$   elements.
Different curves correspond to systems with  different values of
$u$, as shown.}
\label{cloud}
\end{figure}

Figure \ref{cloud}  shows the decay of  $n(t)$ as  a function of
time, for several values of the threshold $u<u_c$. Each curve is
the average of several hundred realizations. We  see that the
decay is monotonous, though for some values  of $u$ the
evolution  is extremely slow.  This happens near $u=0$ and
$u=0.5$, whereas at  intermediate values of the threshold the
decay is faster.

\begin{figure}
\centering \resizebox{\columnwidth}{!}{\includegraphics{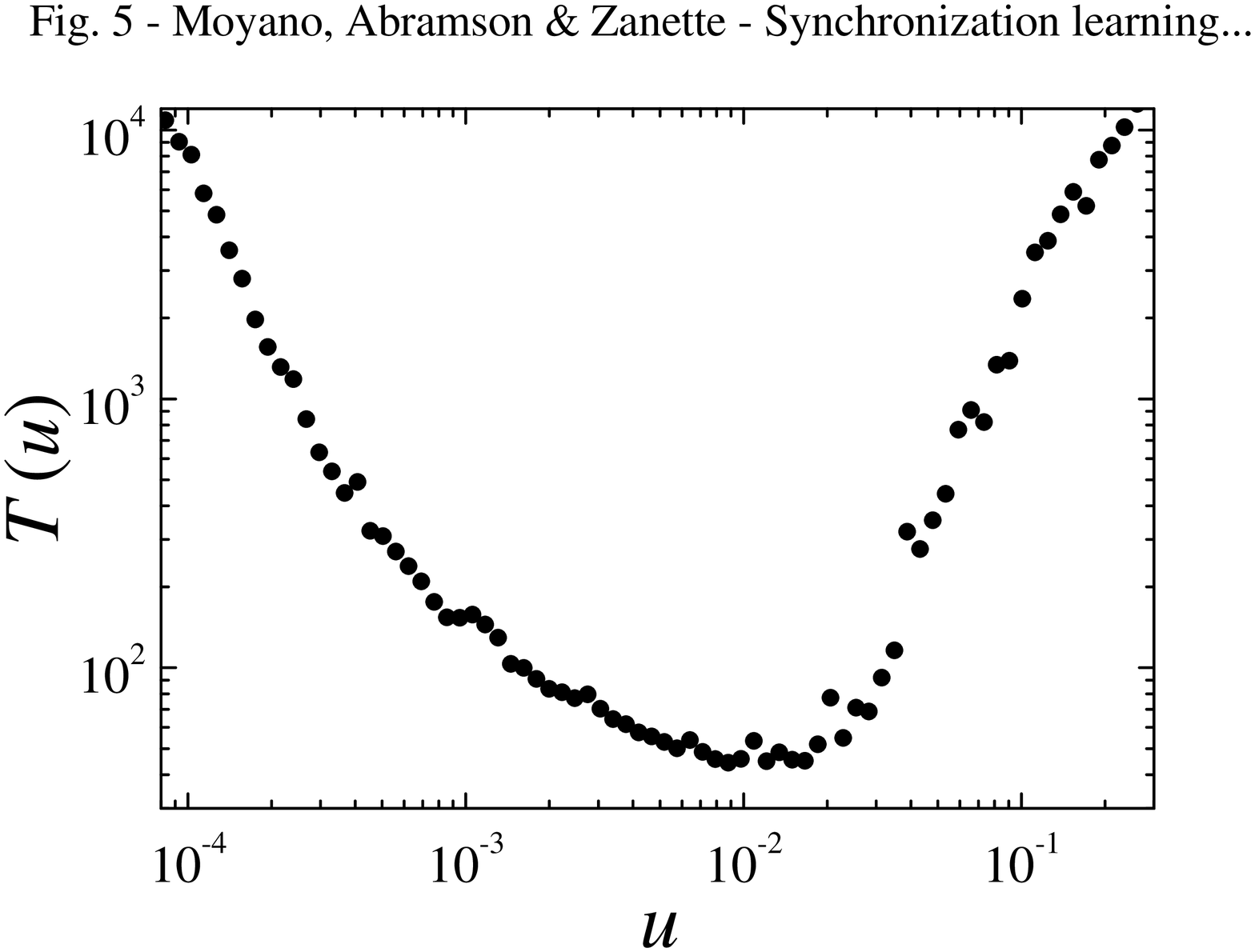}}
\caption{Average learning time $T(u)$ as a function the threshold
$u$ in ensembles of $N=1000$ elements.}
\label{optimal}
\end{figure}

It is apparent from Fig. \ref{cloud} that, for different values of the
threshold,  the functional form of $n(t)$  is not uniform. In order to
define a characteristic time associated  with learning, then, we fix a
reference level $n_0$ and measure the time $T(u)$ needed for $n(t)$ to
reach that level  for  each value of   $u$. A plausible value  for the
reference level   is $n_0 \sim   N^{-1}$,   which indicates  that  for
$t>T(u)$ only a  few elements remain in  the  unsynchronized cloud. In
our simulations,  the learning time  $T(u)$ has been determined taking
$n_0=N^{-1}=10^{-3}$.  It  thus corresponds to  the time  taken by all
elements but one to migrate to the main cluster.  The learning time as
a function of the threshold  $u$ is shown  in Fig. \ref{optimal}. Each
dot stands for the average of several hundred realizations. There is a
clear minimum at $u \approx 0.01$ showing that, as for the performance
of learning, there  is an optimal choice  for the threshold $u$.  Note
that,  as far as $n_0  \sim N^{-1}$, the position  of the minimum does
not depend on the reference level.

The presence of an optimal threshold for our learning algorithm
can be explained as follows.  For large values  of the
threshold,  $u\lesssim 0.5$, the formation of the small  cloud at
$\epsilon \lesssim 0.5$ and the ensuing  appearance   of
intermittent evolution   occur relatively fast. Once this
situation  has been established, however, the elements in  the
cloud spend  very long times in  the ``laminar'' regime and in
unsuccessful  bursts,  whose  amplitude is   not  enough to
drive the elements beyond the threshold. Successful bursts, where
$\Delta x > u$ [see Eq. (\ref{delta})],  become  in fact
increasingly rare  as $u\to u_c$  and, consequently, the learning
time  is expected  to diverge in such limit. At the  other end,
$u   \approx 0$, the threshold  is very narrow   and the elements
keep changing   their coupling constants for long times.   They
need  many   attempts  to  approach  the   average behavior.
Therefore, even  the  initial stage during  which the  main
cluster is formed lasts asymptotically  large times as $u\to 0$.
Since the learning time $T(u)$ should diverge  both at $u=0$ and
at $u=u_c$, it must reach (at  least) a minimum for  an
intermediate value of  the threshold, as fully confirmed by our
numerical results.

\section{Summary and Conclusion}

\label{conclusions}

We have studied an ensemble  of globally coupled  chaotic maps
able to change their individual  couplings in order  to evolve
from an incoherent collective dynamics to  a completely
synchronized state.  The learning procedure  is implemented by
means  of a stochastic    unsupervised algorithm  characterized
by a single parameter   $u$ than measures the stiffness of
learning.  Our numerical results  show that the emergence of
synchronization is only  possible for  a  specific range  of the
parameter   $u$. In fact,   a sharp transition  at  $u_c=0.5$ has
been found, separating  a  regime of successful  learning
($u<u_c$) from a regime  where   the algorithm  fails    to
drive  the  system  to full synchronization ($u>u_c$). In  the
zone of successful learning, in turn, the time needed to reach a
certain level of the learning process has been shown to strongly
depend on $u$. In particular, we have found that  there  is  an
optimal  value, $u\approx 0.01$,   for  which the learning time
is minimum, i.e. learning is fastest.

The   sharp   transition  between    the  regimes  of   successful and
unsuccessful   learning  can be  explained    taking into account  the
intermittent    evolution observed just      before the state of  full
synchronization has  been reached. This kind of   evolution is in fact
typical at the threshold of a synchronization transition \cite{FY}. In
this regime it is  possible   to formulate an  approximate   dynamical
description that  yields the bounds for  the amplitude of intermittent
bursts, during which the elements are subject to the learning process.
These bounds define in turn the maximal value $u_c$ for which learning
is possible.  As for  the optimal value of  $u$,  we have argued  that
both for $u\to u_c$ and for $u\to 0$ the  learning time is expected to
diverge, so that at least one minimum should be found for intermediate
values.  The existence of an optimal value  for the learning stiffness
should be a generic property of a  large class of learning algorithms.
This  point has  already been  discussed  to some extent in connection
with several  training  algorithms for neural  networks \cite{Haykin}.
Indeed,  every  teacher  should    know that  there  is    an  optimal
``pressure'' to  be applied on the  average student to obtain the best
and fastest results in learning.

Several   generalizations to  the   present   model can be   foreseen,
attempting to describe  other situations found in  real systems.   Our
learning   algorithm, in fact, is   based  on  the  comparison of  the
individual state of  each element with a  global quantity, namely, the
average state over  the ensemble. This could  be replaced by a sort of
``local'' criterion,  where the comparison  takes place between pairs
\cite{imit} or small groups of  elements. Moreover, coupling constants
could be subject to smoother  changes, representing a smarter learning
process, instead of the trial-and-error  method used here. In the line
of  some training algorithms for  neural  networks and of optimization
schemes,  a variation of the present  model would  consist in allowing
the learning   stiffness to change with   time.  A suitably controlled
temporal  variation  for $u$ could in  fact   result in  a substantial
decrease   of the learning  times.    Learning itself could be   fully
replaced  by an  evolutionary  mechanism,  in  the spirit  of  genetic
algorithms. In this case, unsuccessful  elements should be  eliminated
and replaced by slightly modified copies of successful elements, which
are the effects   expected   from  natural selection  and    mutation,
respectively. Finally, the individual dynamics of the coupled elements
admits to  be  varied within  an ample class   of behaviors, including
discrete and continuous evolution. The  present work, in summary, is a
first step in  the study of a  wide spectrum of problems,  of interest
from the viewpoint of biology, optimization techniques, and artificial
intelligence.

\end{document}